\documentclass[prb,superscriptaddress,showpacs,twocolumn]{revtex4}

\usepackage{graphicx}
\usepackage[latin1]{inputenc}
\usepackage{amsmath}

\newcommand{\eq}[1]{Eq.~(\ref{#1})}
\newcommand{\fig}[1]{Fig.~\ref{#1}}

\newcommand{\sect}[1]{Sec.~\ref{#1}}

\newcommand{\sij}[1]{\sum_{\langle ij#1\rangle}}
\newcommand{\Na}{NaV$_2$O$_5$}
\newcommand{\gsim}{\raisebox{-3pt}{\,$\stackrel{>}{\sim}$\,}}
\newcommand{\lsim}{\raisebox{-3pt}{\,$\stackrel{<}{\sim}$\,}}
\newcommand{\eV}{{\,\rm eV}}

\begin{document}

\title{Charge order induced by electron-lattice interaction in  \Na}

\author{B. Edegger}
\affiliation{Institut f\"ur Theoretische Physik,
Technische Universit\"at Graz,
A-8010 Graz, Austria}
\affiliation{Institiut f\"ur Theoretische Physik,
Universit\"at Frankfurt, D-60438 Frankfurt, Germany}

\author{H.G. Evertz}
\affiliation{Institut f\"ur Theoretische Physik,
Technische Universit\"at Graz,
A-8010 Graz, Austria}

\author{R.M. Noack}
\affiliation{Fachbereich Physik,
Philipps Universit\"at Marburg,
D-35032 Marburg, Germany}

\begin{abstract}
We present Density Matrix Renormalization Group calculations of
the ground-state properties of quarter-filled ladders including
static electron-lattice coupling. 
Isolated ladders and two coupled ladders are considered, with model
parameters obtained from band-structure calculations for
$\alpha^\prime$-NaV$_2$O$_5$. 
The relevant Holstein coupling to the lattice causes 
static out-of-plane lattice distortions, which appear concurrently 
with a charge-ordered state and which
exhibit the same zigzag pattern observed in experiments. 
The inclusion of electron-lattice coupling 
drastically reduces the critical nearest-neighbor Coulomb repulsion
$V_c$ needed to obtain the charge-ordered state.
No spin gap is present in the ordered phase.
The charge ordering is driven by the Coulomb repulsion
and the electron-lattice interaction.
With electron-lattice interaction, 
coupling two ladders has virtually no effect on $V_c$ or on the
characteristics of the charge-ordered phase.
At $V=0.46\eV$, a value consistent with previous estimates,
the lattice distortion, charge gap, 
charge order parameter, and the effective spin coupling
are in good agreement with experimental data for NaV$_2$O$_5$.
\end{abstract}

\pacs{71.10.Fd, 71.38.-k, 63.22.+m}

\maketitle

\section{Introduction}

Since the observation of a phase
transition at $T_{\rm CO}\approx 34$ K via 
magnetic  susceptibility measurements,\cite{Isobe96} the physical
properties of the low-dimensional inorganic compound NaV$_2$O$_5$
\cite{Lemmens03}
have been investigated intensively. 
The lattice distortions,\cite{Luedecke99}
the opening of a spin gap \cite{Yosihama98} at
$T_{\rm CO}$ or slightly below,\cite{Fagot00,Koeppen98}
and the static charge disproportion $\delta$ between the V ions are
the most remarkable properties of 
this phase transition, which was at first
identified as a spin-Peierls transition like that in CuGeO$_3$.\cite{Hase93}
Later studies\cite{Lemmens03} found
that zigzag charge ordering occurs below the phase transition in \Na,
which is made up of quarter-filled ladders.
It now appears that the opening of the spin gap may
be induced primarily by charge 
transfer\cite{Mostovoy00,Gros99,Grenier01,Gros05} 
in the low temperature 
superlattice.\cite{Grenier02,Sawa02,Smaalen02,Ohwada05,Chitov04a}
%
The discussion on the main driving force of the transition is
still going on.
It was argued recently that the Coulomb repulsion on an isolated
ladder might be too small to cause charge ordering.\cite{Sa00,Gros05}
The inter-ladder coupling, 
which appears to cause the spin gap, 
would then also have to be responsible for the charge ordering. 
However, the electron-lattice coupling has been found to contribute to
the charge-ordering 
transition.\cite{Sherman99,Fischer99,Riera99,Bernert02,Clay03,Aichhorn04} 

In this work, we concentrate on the effects of electron-lattice interactions 
on the ground state of NaV$_2$O$_5$.
Due to the asymmetric  crystal environment of the V ions in \Na,
the strongest lattice coupling is a 
Holstein-like electron-phonon interaction
involving the $d_{xy}$ electrons.\cite{Sherman99,Spitaler04}
We investigate the extended Hubbard model with Holstein coupling,
introduced in Ref.~\onlinecite{Aichhorn04},
using high precision Density Matrix Renormalization Group (DMRG)
calculations on isolated and on two coupled ladders.
We show that, indeed, including the electron-lattice coupling causes
charge ordering to appear 
on isolated as well as on two coupled ladders
at significantly smaller Coulomb repulsions,
with values which are consistent with most independent estimates.

In \sect{sec2} we present the model and discuss the parameter values
and observables.
We discuss results for an isolated ladder in \sect{1ladder}
as a function of the nearest-neighbor Coulomb repulsion. 
Section \sect{2ladder} treats coupled ladders,
and is followed by our conclusions.

\begin{figure}[t]
  \centering
  \includegraphics[width=0.25\textwidth]{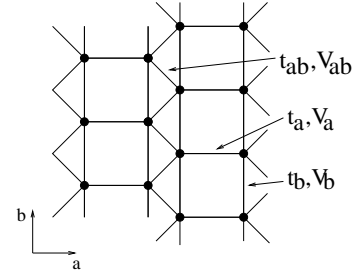}
  \caption{\label{fig:ladders}%
Schematic depiction of the vanadium ladders in NaV$_2$O$_5$. 
The three types of hopping matrix
elements ($t_a, t_b, t_{ab}$) and nearest-neighbor Coulomb repulsion
($V_a, V_b, V_{ab}$) are indicated. 
Two ladders are shown.}
\end{figure}

\section{Model}
\label{sec2}
NaV$_2$O$_5$ is the only known quarter-filled ladder
compound.\cite{Smolinski98}
Ladder-like structures are  formed by the vanadium ions
and are only weakly coupled. 
A useful microscopic description is provided by an extended Hubbard
model (EHM) with Coulomb repulsion between nearest-neighbor sites.
The properties of the EHM without electron-lattice coupling were
studied using 
mean field approaches,\cite{Seo98,Thalmeier98,Cuoco99,Mostovoy00}
perturbation theory,\cite{Yushankhai01}
the Dynamical Mean-Field Theory,\cite{Mazurenko02} exact 
diagonalization,\cite{Cuoco99,Huebsch01,Huebsch01a,Aichhorn02,Aichhorn04}
bosonization,\cite{Orignac03} Quantum Monte Carlo,\cite{Gabriel05} 
and Cluster Perturbation Theory.\cite{Aichhorn04a}
Detailed studies of the EHM were also performed using the
DMRG,\cite{Vojta99,Vojta01} but not at specific values of the
coupling parameters appropriate for \Na. 

A complete microscopic description of NaV$_2$O$_5$
must incorporate the lattice distortion 
observed by x-ray diffraction in the low
temperature phase.\cite{Luedecke99}
The distortions of the vanadium ions
considerably change the distance to the neighboring oxygen atoms
directly above or below the vanadium site.\cite{Luedecke99}
Recent calculations using Density Functional Theory in the Local
Density Approximation (LDA) have shown that the main effect
of these distortions is a change of the local potentials,
i.e., a Holstein-like coupling $H_{e-l}$.\cite{Spitaler04} 
Indeed, the corresponding phonons are the strongest-coupling $A_g$
modes\cite{Spitaler04} 
in \Na; in the current study, we restrict the electron-lattice coupling to
these distortions.
The contribution of the lattice deformation to the Hamiltonian
can be approximated by a parabolic potential $H_l$.
This yields the microscopic Hamiltonian proposed 
and studied on small systems in Ref.~\onlinecite{Aichhorn04},  
\begin{equation}\label{ham}
    H=H_{\rm EHM}+H_{ l}+H_{ e-l}
\end{equation}
with
\begin{subequations}
\begin{align}
    H_{\rm EHM}=&-\sum_{\langle ij\rangle,\sigma}t_{ij}\left(c_{i\sigma}^\dagger
    c_{j\sigma}^{\phantom{\dagger}}+\mbox{h.c.}\right)\label{hehm}\nonumber\\
    &+U\sum_in_{i\uparrow}n_{i\downarrow}+\sij{}V_{ij}n_in_j,\\
    H_{l}= &\kappa\sum_i\frac{z_i^2}{2},\label{eq_2b}\\
    H_{e-l}=&-C\sum_iz_in_i \; , \label{eq_2c}
\end{align}
\end{subequations}
where $n_i=n_{i\downarrow}+n_{i\uparrow}$ is the occupation number 
and $z_i$ the distortion at site $i$
(in units of 0.05 \AA).
The Hamiltonian contains
the effective lattice force constant $\kappa$ and a large Holstein
constant $C$.
The placement of the hopping terms $t_{ij}$ and the Coulomb repulsion
$V_{ij}$ in the lattice structure are depicted in \fig{fig:ladders}. 
The parameters are taken from LDA calculations\cite{Spitaler04};
the hopping amplitudes $t_{ij}$ are compatible with earlier calculations.
\cite{Smolinski98,Horsch98}
The lattice parameters $C$ and $\kappa$ were extracted 
by comparing the total energy and the
inter-ionic forces in distorted and undistorted lattices. 
They correspond to the V-O-stretching perpendicular to the ladder plane
observed in the $970 \, \mbox{cm}^{-1}$ phonon mode. 
As in Ref.\ \onlinecite{Aichhorn04}, 
we take
\begin{align}
 t_a   &= 0.35\eV = 2 \,t_b   \, , \quad
                 t_{ab}=0.17 \,t_a ~\text{and}~ 0.33\, t_a\, , \nonumber\\
 V_a   &= V_b = V_{ab} = V \, ,   \qquad\qquad\,       U=8.0\, t_a \,
                 , \nonumber\\
 \kappa&=0.125\, t_a \, , 
\qquad\qquad \mbox{and}\;\;\;  C=0.35\, t_a \, .
\end{align}
We compare results for two different values, $0.17\,t_a$ and $0.33\,t_a$,
of the interladder hopping because
existing estimates of these parameters 
differ.\cite{Spitaler04,Smolinski98,Horsch98}
%
The Coulomb repulsion is very difficult to compute;
estimates in the literature vary strongly.%
\cite{Cuoco99,Sa00,Yaresko00,Yushankhai01,Mazurenko02,Mostovoy00}
In our study, we take $V_a=V_b=V_{ab}=V$ for simplicity,
and investigate the model as a function of $V$.


We have performed DMRG calculations on isolated ladders of length of up to
$L=80$ and on two coupled ladders of length of up to $L=24$,
applying open boundary conditions along the chains ($b$-direction).
In the case of two coupled ladders, we have taken periodic boundary
conditions in the $a$ direction.

We have measured the  charge order parameter $m_{\rm CO}$, defined by
\begin{align}
    m_{\rm CO}^2=\frac{1}{N^2\langle n\rangle^2}\sum_{ij}e^{i{\mathbf
    Q}({\mathbf R}_i-{\mathbf R}_j)}\left(\langle n_i n_j\rangle - 
       \langle n\rangle^2 \right) \, ,
\end{align}
where ${\mathbf Q}=(\pi,\pi)$ and $N$ is the total number of sites in the
system. 
The occupation number of 
the two inequivalent V ions is then $(1 \pm m_{\rm CO})/2$.

The charge gap $\Delta_C(L)$ and the spin gap $\Delta_S(L)$ are determined 
using 
\begin{align}
\Delta_C(L) &= \frac{1}{2}[E_0(L,N+2)+E_0(L,N-2)-2 E_0(L,N)] \nonumber\\ 
\Delta_S(L) &= E_0(L,N,S_z=1)-E_0(L,N,S_z=0) \; ,
\label{gaps_gap}
\end{align}
where $E_0(L,N)$ is the ground state energy of the system with $L$
rungs per ladder and $N$ electrons.\cite{Vojta01}
Results are extrapolated to $L=\infty$ 
using linear and quadratic fits in $1/L$ and by applying finite-size scaling.

\section{Isolated ladder}
\label{1ladder}
In this section, we present DMRG calculations on isolated
quarter-filled ladders ($V_{ab}=0$, $t_{ab}=0$). We consider systems
with ($C=0.35$) and without ($C=0$) electron-lattice interactions
and compare the results. 

%
\begin{figure}[tb] 
  \centering
\includegraphics[width=0.48\textwidth]{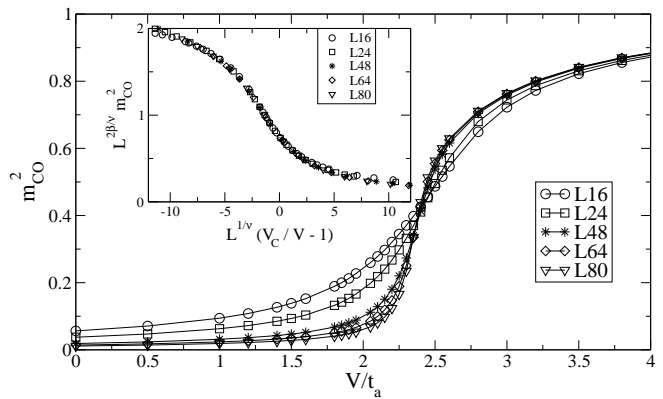}
  \caption{\label{fig:orderpar1L_nodist}%
    Square of the charge order parameter, $m_{\rm CO}^2$, 
    on an isolated ladder without coupling to the lattice,
    as a function of the nearest neighbor Coulomb repulsion $V$
    for several ladder lengths  $L$.
    The inset shows the finite size scaling data collapse,
    with exponents $\nu=1$, $\beta=\frac{1}{8}$, and critical Coulomb repulsion $V_C=2.31(2)$.
   }
\end{figure}
Isolated quarter-filled ladders
without coupling to the lattice were investigated by 
Vojta et.\ al.\cite{Vojta99,Vojta01} for a range of parameters.
The parameters studied did not, however, include the couplings
relevant for \Na, 
which is characterized by large hopping along the rungs, $t_a\simeq 2\,t_b$,
large Coulomb repulsion $U\simeq 8 \,t_a$, and charge ordering.
Ref.~\onlinecite{Vojta01} found qualitatively different behavior
for $t_a \lsim \,t_b$, where the spin gap was found to be finite,
and $t_a \gsim \,t_b$, where the spin gap was found to vanish.

In \fig{fig:orderpar1L_nodist}, we display 
the square of the charge order parameter
$m_{\rm CO}^2$ as a function of the nearest-neighbor Coulomb repulsion $V$
for different system sizes. 
At a critical value $V_c$ above $2\,t_a$,
there is a quantum phase transition to a phase with finite charge order.
The essential features of this transition can be described by a model
with a single charge degree of freedom on each rung, i.e., a pseudospin. 
When the hopping between rungs is neglected, 
one arrives at the Ising model in a transverse field 
(IMTF),\cite{Mostovoy00,Aichhorn02}
which can be solved exactly.\cite{Lieb61}
Indeed, a finite-size scaling analysis for $m^2_{CO}$ in which
$L^{2\beta/\nu} m^2_{CO}$ is plotted as a function of
$L^{1/\nu}(\frac{V_c}{V}-1)$ with the critical exponents
$\nu=1$,  $\beta=\frac{1}{8}$ of the IMTF, 
shown in the inset, 
collapses all data points onto a universal curve and yields a critical
Coulomb repulsion of $V_C=2.31(2)\,t_a$.
We have also performed a similar scaling analysis 
without fixing $\nu$ and $\beta$, and obtain a reasonable data
collapse with $\nu=1.0(1)$, $\beta=0.125(20)$, and $V_C=2.31(4)$.
We note that $\beta=\frac{1}{8}$ (which is also the exponent of the 2d
Ising model) is fairly close to the value extracted from experimental results 
for \Na,\cite{Fagot00,CritExp}, about 0.15--0.20.
However, the effective Coulomb coupling $V$ in \Na\ is probably
considerably below 
$V_C\simeq 2.31\,t_a$.\cite{Cuoco99,Sa00,Yaresko00,Yushankhai01,Mazurenko02,Mostovoy00}
Therefore, the Coulomb repulsion alone is 
insufficient to cause the charge ordering in \Na.

\begin{figure}[tb] 
  \centering
\includegraphics[width=0.48\textwidth]{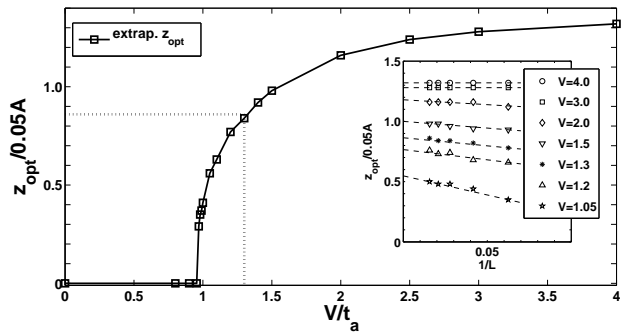}
  \caption{\label{fig:distopt1L}
    Optimal zigzag distortion $z_{opt}$,
    extrapolated to $L=\infty$, as a  function of $V$ 
    on an isolated ladder system with electron-lattice coupling. 
    The inset shows the linear 1/L extrapolation for some values of $V$.
    The experimentally determined distortion in \Na\ corresponds to $V^* = 1.3 \,t_a$,
    indicated by the dotted lines.
  }
\end{figure}
When the electron-lattice coupling is taken into account,
the behavior of the charge order parameter changes drastically,
as we will now show.
The ground state energy $E_0(\{z_i\})$ of the
Hamiltonian $H$ (\eq{ham}) is a function of the independent
classical lattice distortions $z_i$ on each site.
First, we have determined the optimal distortion pattern on ladders of
up to 16 sites using classical Monte Carlo simulations carried out in
the space of all $z_i$.
 
In these simulations, the ground state energies are determined 
by exact diagonalization. 
We use parallel tempering \cite{Marinari96} to find all relevant
distortion patterns.
We find that at low $V$ the optimum lattice distortion 
(lowest total energy)
is $z_i=0$,
i.e., no distortion. 
Above a critical value $V_c \approx 0.95 \, t_a$,
there are two degenerate optimal configurations,
with finite zigzag lattice distortions
\begin{equation}\label{eq_4}
    z_i=z e^{i{\mathbf Q}\cdot{\mathbf R}_i},\quad {\mathbf
    Q}=(\pi,\pi) \; ,
\end{equation}
where ${\mathbf R}_i$ labels the lattice sites.
The zigzag pattern of the lattice distortions agrees with 
the experimentally observed pattern.\cite{Luedecke99}
In the following, we therefore assume a zigzag pattern.
We determine the ground state by minimizing $E_0(z)$ as a function of 
the lattice distortion $z$, similarly to
Ref.~\onlinecite{Aichhorn04}.
The position of the minimum defines the optimal distortion $z_{opt}$.

The optimal distortion $z_{opt}$ 
and the extrapolation to the thermodynamic limit are presented in
\fig{fig:distopt1L}.
The distortion becomes finite at a critical
Coulomb repulsion $V_C=0.95(1)$.
From \fig{fig:distopt1L} and from the experimentally determined size of the
zigzag-distortion,\cite{Luedecke99}
($z_{exp} \, \approx \, 0.85 \;\times\;  0.5\AA$),
we  obtain an estimate for the effective Coulomb repulsion in \Na\ 
of $V^* \!=\! 1.3\, t_a$, well within the range of most earlier 
estimates\cite{Cuoco99,Sa00,Yaresko00,Yushankhai01,Mazurenko02}.

\begin{figure}[tb] 
  \centering
\includegraphics[width=0.48\textwidth]{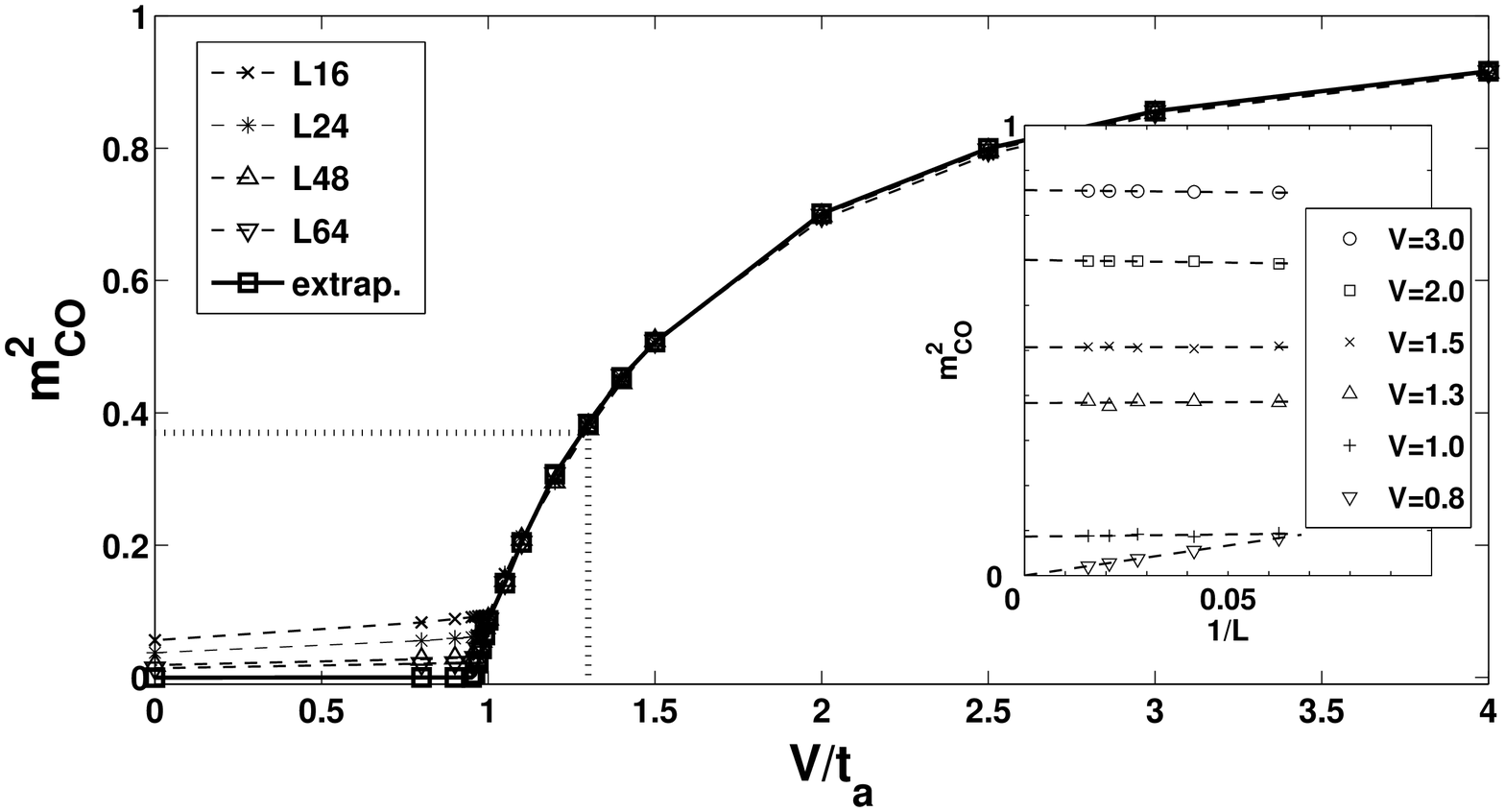}
  \caption{\label{fig:orderpar1L_dist}%
    Square of the charge order parameter, $m_{\rm CO}^2$,
    calculated at the optimal distortion $z_{opt}$
    on an isolated ladder with electron-lattice coupling 
    as a function of $V$ and for several ladder lengths $L$.
    The solid line is the result of a linear $1/L$ extrapolation,
    illustrated in the inset for selected values of $V$.
    The dotted lines mark $V^*=1.3t_a$.
   }
\end{figure}
The square of the charge order parameter $m_{\rm CO}^2(L)$,
calculated at the optimal distortions $z_{opt}(L)$, is shown
in \fig{fig:orderpar1L_dist}, along with a finite-size extrapolation.
Order sets in at the same $V$ as the lattice distortion $z_{opt}$.
For the $1/L$-extrapolated values, 
$m_{CO}^2$ is proportional to the square of the optimal
distortion $z_{opt}^2$ 
at all $V$.
In a finite-size scaling analysis (not shown), 
the scaling of $m_{CO}$ is consistent with the mean-field exponent
$\beta=\frac{1}{2}$, but not with the IMTF exponent
$\beta=\frac{1}{8}$.

A comparison of \fig{fig:orderpar1L_nodist}
and \fig{fig:orderpar1L_dist}  illustrates the
substantial 
decrease of the critical Coulomb repulsion 
due to the electron-lattice coupling. 
The critical value $V_c$ 
decreases from $V_c = 2.31(2)\,t_a$ without electron-lattice coupling
to $V_c =  0.95(1) \,t_a$ with electron-lattice coupling. 
The critical exponent of $m_{CO}$ is clearly smaller than unity in both cases,
but changes from $\beta\simeq\frac{1}{8}$ to $\beta\approx\frac{1}{2}$
when the electron-lattice coupling is switched on.
At the coupling $V^*$ 
at which the lattice distortion matches the experiment value, 
we find $m^2_{CO}\approx 0.37$, 
very close to the experimental results of about 0.35\cite{Grenier01}
and 0.37.\cite{Sawa02}

\begin{figure}[tb] 
 \centering
 \includegraphics*[width=0.49\textwidth]{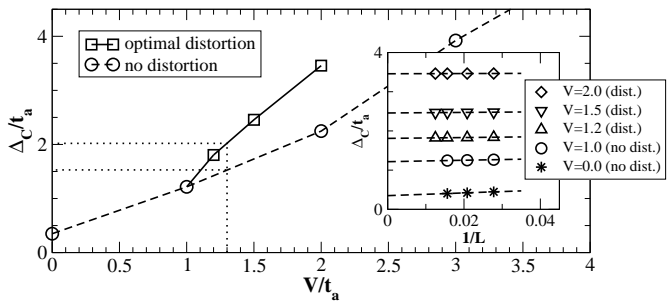}
 \caption{\label{fig:charge_gap}%
 Charge gap $\Delta_C$ on an isolated ladder as a function of $V$,
 with (solid) and without (dashed) electron-lattice coupling. 
 The results at $V^*=1.3 \, t_a$ are indicated by dotted lines. 
 Inset: linear finite-size extrapolation in $1/L$
 for systems with (dist.) and without (no dist.) lattice distortion.   
}
\end{figure}
The charge gap $\Delta_C$
is also influenced by the presence of electron-lattice coupling,
as shown in \fig{fig:charge_gap}.
The charge gap increases with
increasing $V$ and lattice distortion
and does not vanish\cite{Vojta01} at $V=0$. 
%
At $V^*=1.3\,t_a$, the charge gap agrees reasonably well with the
experimental value for the optical gap in \Na.\cite{Presura00}

\begin{figure}[b] 
 \centering
\includegraphics*[width=0.5\textwidth]{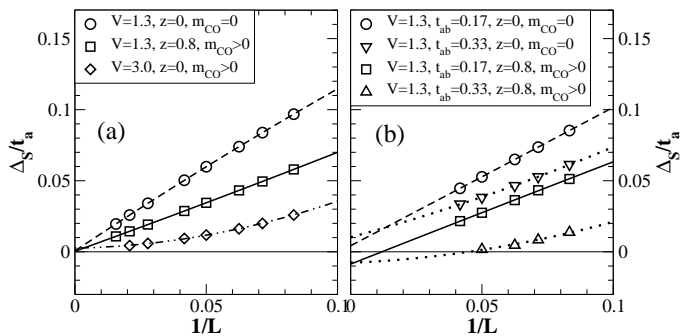}
 \caption{\label{fig:SGs}%
 Spin gap $\Delta_S$ for (a) an isolated ladder and (b) two coupled ladders.
 Lines represent  a quadratic fit in $1/L$.
 The boxed descriptions of the curves are in the same vertical order
 as the data.
 }
\end{figure}

The spin gap of an isolated quarter-filled ladder 
without electron-lattice coupling
has been shown by Vojta {\it et al.}\cite{Vojta01} to vanish at moderate $V$
when the rung hopping $t_a$ is larger than 
$t_b$, as is the case in \Na.
In \fig{fig:SGs}(a), we show some representative results
of our calculations of the spin gap,
with and without electron-lattice coupling, 
together with a quadratic $1/L$-extrapolation.
We find that the {\em spin gap extrapolates to zero 
on an isolated ladder at all $V$} in both cases.
Charge ordering on an isolated ladder is therefore not sufficient to
induce a spin gap.

\begin{figure}[t]  
 \centering
 \includegraphics*[width=0.35\textwidth]{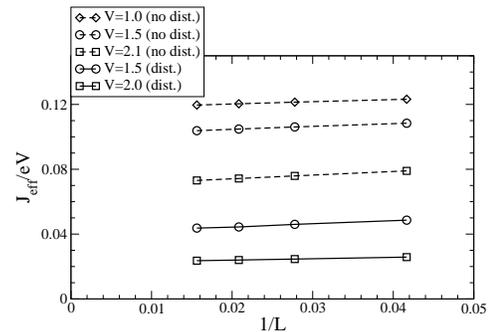}
 \caption{\label{fig:detJ}%
Effective magnetic exchange coupling 
$J_{\text{eff}}$ (\eq{eq_Jeff})
as a function of inverse lattice size $1/L$,
for different values of $V$,
with and without lattice distortion.
 }
\end{figure}
\begin{figure}[b] 
 \centering
 \includegraphics*[width=0.45\textwidth]{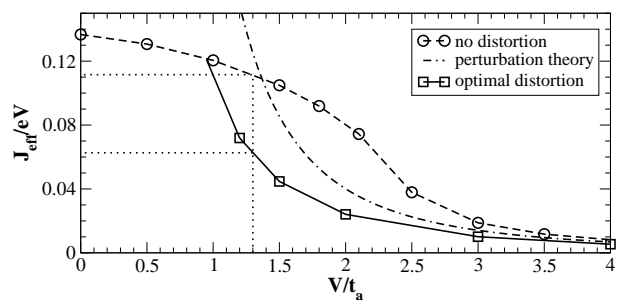}
 \caption{\label{fig:Jeff_1L}%
 Effective magnetic exchange coupling $J_{\text{eff}}$
 on an isolated ladder as a function of $V$, with (solid line) and
 without (dashed line)
 electron-lattice coupling. 
 The dashed-dotted line is the result of perturbation theory \cite{Vojta01}
 for the undistorted lattice,
 valid only at large charge order.
 }
\end{figure}

The magnetic behavior of the quarter-filled Hubbard ladder can be approximated
by an antiferromagnetic Heisenberg model.\cite{Smolinski98,Horsch98,Vojta01}
We determine an effective magnetic exchange interaction $J_{\text{eff}}$
by equating the finite-size spin gap in our model to that of a
Heisenberg chain
with exchange constant $J$, i.e.,
\begin{equation}\label{eq_Jeff}
 \frac{\Delta_S^\text{Heisenberg}(L)}{J} ~~=~~  
\frac{\Delta_S^\text{Hubbard}(L)}{J_{\text{eff}}(L)} \; .
\end{equation}
The results are plotted in \fig{fig:detJ} as a function of inverse
system size $1/L$.
There is only a weak $L$-dependence, and the scaling is linear in $1/L$.
Previous work using exact diagonalization\cite{Aichhorn04} showed that 
the behavior of the dynamical spin correlations  $S(k,\omega)$ for the
EHM with electron-lattice coupling 
also closely resemble those of the Heisenberg chain.
Remarkably, our results for $J_{\text{eff}}$ agree very well with 
the estimates from the spin dispersion in Ref.~\onlinecite{Aichhorn04}.

The resulting values of $J_{\text{eff}}$ are plotted in
\fig{fig:Jeff_1L} as a function of $V$.
As $V$ is increased,  the charge order parameter increases,
and $J_{\text{eff}}$ decreases.
This behavior is in accordance with the experimental observation that
$J_{\text{eff}}$ becomes smaller at lower temperature, where charge
order increases.
Part of the reduction in $J_{\text{eff}}$ is due directly to the
charge occupation of neighboring sites along the chains, which implies
a reduction by a  factor of $(1-m^2_{CO})$.\cite{Gros99}
Our results, however, are far from such a simple quadratic dependence
on $m^2_{CO}$.
%

\section{Coupled ladders}
\label{2ladder}
%
\begin{figure}[tb]  
 \centering
 \includegraphics*[width=0.45\textwidth]{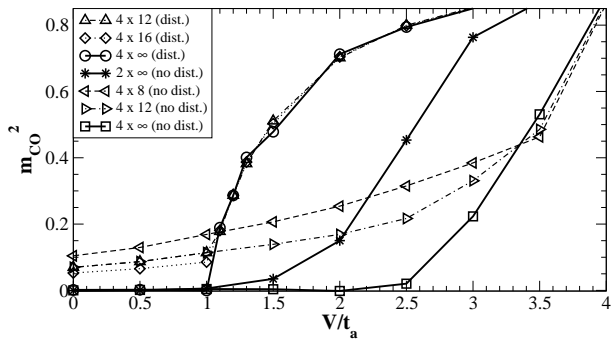}
 \caption{\label{fig:mCO_2L}%
 Square of the charge order parameter, $m_{\rm CO}^2$, for two coupled ladders
 ($4 \times L$, $t_{ab}=0.33\,t_a$) 
 with  periodic boundary conditions in the $a$ direction
 and with and without lattice distortions. 
 The results for infinite length are obtained using a linear
 extrapolation in $1/L$.
 Also included for comparison are similarly extrapolated results for
 an isolated ladder ($2\times \infty$).
 }
\end{figure}

The occurrence of a spin gap in \Na\ 
may largely be due to the coupling of ladders.\cite{Mostovoy00,Gros99,Grenier01,Gros05} 
We have therefore studied a system of two coupled ladders
with periodic boundary condition in the $a$ direction.
We have determined the pattern of optimal lattice distortions in the
same way as for the isolated ladder (for $t_{ab}\,=\,0.33\,t_a$),
using exact diagonalization on a 16-site system,
and find a simultaneous zigzag distortion on both ladders to be optimal.
Due to the lattice structure (see \fig{fig:ladders}),
the energy is invariant under a shift of the distortion by one rung on
either ladder.
Using the DMRG, we have then examined 
larger systems.
We have determined the optimal distortion $z_{opt}$
from the minimum of the ground state energy 
$E_0(z)$ as a function of the uniform zigzag distortion.
The result is very similar to the lattice distortion in the isolated
ladder at the same $V$. 
The effective Coulomb repulsion determined from the experimental
lattice distortion is therefore still $V^*=1.3\,t_a$. 

Without electron-lattice coupling, convergence of the DMRG calculations is
difficult to achieve, restricting the largest length to at most $L=12$ at
large $V$.
The three right-most curves in \fig{fig:mCO_2L} show the results 
for the charge order parameter as a function of $V$ for lattice sizes
$L=8$, $L=12$, and for a  $1/L$ extrapolation. 
For comparison, the results for a single ladder without
electron-lattice coupling extrapolated in the same way
are also shown (middle solid line marked with stars).
Clearly, the system consisting of two coupled ladders tends to order
at even larger values of $V$ than the isolated ladder,
leading to an even larger discrepancy with the estimates for $V$ in \Na.

At the lattice sizes we were able to reach,
the effects of the open boundaries, especially effects caused by the 
different number of neighbors at the boundary, are considerable.
While varying the boundaries, e.g., compensating for missing neighbors,
did affect the results shown in \fig{fig:mCO_2L}, 
it did not change the overall tendency to shift the phase transition to
larger $V$.

For systems with electron-lattice coupling, we were able to reach
larger sizes, up to $4\times 16$ sites at large $V$ and up to 
$4 \times 24$ at $V^*=1.3 \,t_a$.
The results, as can be seen in \fig{fig:mCO_2L}, 
are almost identical to the results on a single ladder of the same
length $L$, and are almost independent of length in the ordered phase.
Charge ordering still takes place at $V\simeq 0.95 \,t_a$.

The spin gap of the coupled ladder system is displayed in 
\fig{fig:SGs}(b) for $V^*=1.3\,t_a$.
Since estimates of the inter-ladder hopping $t_{ab}$ in \Na\ vary,
\cite{Spitaler04,Smolinski98,Horsch98}
we have performed our calculations for two values, $t_{ab}=0.17 \,t_a$
and $0.33 \,t_a$.
At $t_{ab}=0.17 \,t_a$, the spin gap for two coupled ladders both with
and without lattice distortion is
similar in value and finite-size behavior to that for the isolated
ladder,
which vanishes in the thermodynamic limit. 
When the inter-ladder hopping is increased to $t_{ab}=0.33 \,t_a$, 
the spin gap becomes smaller and its finite-size dependence appears to
change in both cases.
The physics of the individual ladders thus appears to be 
noticeably affected by inter-ladder coupling at this larger hopping.

Extrapolations of the gaps 
using a fit to a quadratic polynomial in $1/L$ are indicated by the
lines in the figure. 
The value of the spin gap in the $L\to \infty$ limit for all parameter
values is approximately zero, $|\Delta_S(L=\infty)|<0.01\,t_a$. 
Some of the extrapolations in \fig{fig:SGs} yield values that are
slightly negative.
The magnitude of these negative values gives
as a minimal error estimate for the extrapolation. 
Since the magnitudes of the positive extrapolations 
are of similar size, we conclude that the extrapolated spin gaps are
zero to within errors.
In addition, the small extrapolated gaps are significantly
smaller than the spin gap found experimentally in NaV$_2$O$_5$,
$\Delta_{S} \approx 10\,{\rm meV} \approx 0.03\,t_a$.
The spin gaps at other values of $V$ ($V=0.9\,t_a,\, 1.5\,t_a$) 
behave similarly.

We conclude that when two ladders are coupled, there is no significant
indication that a spin gap opens. 
This result is not unexpected, however, because the mechanism which
has been proposed as the cause for the spin gap in the charge-ordered
system,%
\cite{Mostovoy00,Gros99,Grenier01,Gros05} 
namely an in-ladder dimerization of effective spin couplings,
requires a charge-ordering pattern with a period of {\em four} ladders in
the $a$ direction; such a pattern is observed 
in \Na.\cite{Grenier02,Sawa02,Smaalen02,Ohwada05}

\section{Conclusions}
\label{conclusions}

\begin{table}[tb]
\vbox{
 \begin{tabular}{||c||c|c||}
  \hline\hline
                    & EHM+lattice     &      Experimental\\
                    & $V=1.3\,t_a$    &      data\\
  \hline \hline
  Distortion/0.05\AA& $ 0.84$  & $ 0.85$, Ref.~\onlinecite{Luedecke99}\\
  $m_{\rm CO}^2$     & $0.37$   & $0.35-0.37$, Refs.~\onlinecite{Grenier01,Sawa02}\\
  $J_{\text{eff}}$         & $ 63\,{\rm meV}$& $ 60\,{\rm meV}$, Ref.~\onlinecite{Grenier01}\\
  $\Delta_C$        & $\,0.7 \eV$&$ 0.9 \eV$, Ref.~\onlinecite{Presura00}\\
  \hline\hline
 \end{tabular}
 \caption{\label{compare_data}%
  Comparison of experimental data for \Na\ 
  to the results for the Extended Hubbard Model with lattice coupling,
  at $V^*=1.3\,t_a$.}
}
\end{table}

In this paper, we have investigated the influence of Coulomb repulsion
and electron-lattice interactions in quarter-filled
ladder materials such as \Na\ by treating the extended Hubbard model
with and without 
%
coupling to static lattice distortions in the $c$ direction. 
Our calculations show that the electron-lattice coupling
drastically affects the physical properties of the low-temperature phase. 
It causes lattice distortions to appear concurrently with
charge ordering and with the same spatial zigzag pattern. 
The transition to the charge-ordered phase
is shifted to a much smaller critical value of the nearest-neighbor Coulomb
repulsion ($V_c \approx 0.95\,t_a$). 
The charge gap is increased and the effective magnetic exchange
is reduced by the electron-lattice coupling.
These results remain unchanged when two ladders are coupled.
The spin gap extrapolates to zero in all cases.

When the electron-lattice coupling is included, the
properties of both the isolated ladder and two coupled 
ladders  are in remarkably good agreement with experimental
data for \Na,
at an effective Coulomb repulsion $V^*=1.3\,t_a$ determined
by matching the experimental lattice distortion.
This Coulomb repulsion is well within the range 
previously estimated for \Na.
As shown in Table~\ref{compare_data}, 
we find simultaneous
agreement for the 
zigzag distortion, the size of the charge order parameter, the
effective spin interaction, 
and for the charge gap with experimental data.

While results for the isolated ladder without
electron-lattice coupling would also be reasonably consistent with the
experimental values, 
an unrealistically large value for the Coulomb repulsion, 
$V\simeq 2.5 \,t_a$, would be required, and the experimentally
observed lattice distortion would be not present.
The value of the Coulomb repulsion necessary for such an agreement
becomes still larger when two ladders without electron-lattice
interaction are coupled.

We conclude that an interplay of charge ordering
and lattice distortion can drive the phase transition to a
charge-ordered state in NaV$_2$O$_5$, independent of the occurrence of
a spin gap.
The explanation of the spin gap will likely require a unit cell with
at least four coupled ladders.
Corresponding work is in progress.\\

\begin{acknowledgments}
This work has been supported by the Austrian Science Fund FWF, project P15520.
The authors would like to thank M.~Aichhorn, C.~Gros, T.C.~Lang, F.~Michel, and E. Sherman
for stimulating discussions.
\end{acknowledgments}


\end{document}